\documentclass[aps,prl,superscriptaddress,showpacs,twocolumn]{revtex4}
\usepackage{graphicx,amsmath, amssymb, pifont}
\usepackage[usenames]{color}

\begin{document}

\title{Effect of Microwaves on the Current-Phase-Relation in SNS Josephson Junctions}

\author{M.~Fuechsle, J.~Bentner, D.A.~Ryndyk, M.~Reinwald, W.~Wegscheider, and C.~Strunk}

\begin{abstract}
We investigate the current-phase-relation (CPR) of long diffusive
superconductor-normal metal-superconductor (SNS) Josephson junctions
in thermodynamic equilibrium and under microwave irradiation. While
in equilibrium good agreement with the predictions of
quasi\-classical theory is found, we observe that the shape of the
CPR can be strongly affected by microwave irradiation. Close to a
Josephson-phase difference $\varphi\simeq\pi$, the supercurrent can
be strongly suppressed when increasing the rf-power. Our results can
be understood in terms of microwave excitation of low-lying Andreev
bound states across the mini-gap in the junction. In the frequency
interval studied, this mechanism becomes important, when the
mini-gap closes at $\varphi\simeq\pi$.
\end{abstract}
 \maketitle

The fundamental mechanism of Cooper-pair transfer across
superconductor-normal metal-superconduc\-tor (SNS) Josephson
junctions relies on the existence of Andreev bound states (ABS) in
the normal metal \cite{kulik}. The energies $\epsilon$ of the ABS
depend on the phase difference $\varphi$ between the two
superconductors. Diffusive normal metals exhibit a continuous
spectrum of ABS, described by a complex supercurrent spectral
density $j_\mathrm{S}(\epsilon,\varphi)$  \cite{ilichev,heikk}.
Assuming perfectly transparent interfaces at a distance $d$,
$j_\mathrm{S}(\epsilon,\varphi)$ is determined by two parameters,
namely the energy gap $\Delta$ of the superconductor, and the
Thouless energy $\epsilon_\text{th}=\hbar D/d^2$, where $D$ is the
diffusion constant. The ABS spectrum exhibits a minigap
$\epsilon_\text{g}(\varphi)$, which closes as $\varphi$ approaches
$\pi$. The total supercurrent is given by \cite{ilichev}
\begin{equation}\label{SC_spect}
I_\mathrm{S}(\varphi)=\frac{1}{eR_\mathrm{N}}\int_0^\infty {\rm
Im}\,j_\mathrm{S}(\epsilon,\varphi)
\left[1-2f(\epsilon)\right]d\epsilon,
\end{equation}
where $R_\mathrm{N}$ is the resistance of the normal conductor and
in thermal equilibrium $f(\epsilon)$ is the Fermi distribution
function determining the occupation probability of the ABS at
temperature $T$. As opposed to the case of conventional tunneling
junctions, for ideal NS interfaces higher order scattering processes
become important. These processes correspond to the phase-coherent
transfer of multiple Cooper pairs across the junction, and manifest
themselves in a non-trivial Fourier representation
$I_\mathrm{S}(T,\varphi)=\sum_jI_\mathrm{C}^{(j)}(T)\sin (j\varphi)$
of Eq.~1. In thermodynamic equilibrium, the higher harmonics
$I_\mathrm{C}^{(j)}$ are rapidly suppressed for temperatures larger
than $\epsilon_\text{th}/k_\mathrm{B}$.

Attempts to detect higher harmonics in the CPR of diffusive
SNS-junctions by measuring the magnetic response of loops with an
embedded junction date back to the 70's \cite{waldram}, but failed
because of the contradictory requirements of large
$\epsilon_\text{th}$ (very low temperatures and short junctions) and
low $I_\mathrm{C}$ (higher $T$ and long junctions) to avoid
hysteretic switching of the loop \cite{ilichev}. As an alternative
method, the detection of subharmonic Shapiro steps was suggested
\cite{lehnert,dubos}. These experiments were successful, but
resulted in a very surprising non-monotonic $T$-dependence of the
higher order coefficients $I_\mathrm{C}^{(j)}$. As opposed to the
monotonic exponential suppression with $T$ expected from theory
\cite{heikk}, the subharmonic Shapiro steps were observed most
clearly at $T> T_\text{th}$. At the highest temperatures even
$I_\mathrm{C}^{(1)}<I_\mathrm{C}^{(2)}$ and
$I_\mathrm{C}^{(1)}<I_\mathrm{C}^{(3)}$ were found \cite{dubos}.
These findings have so far remained in contradiction with the
equilibrium quasi-classical theory and were explained by the
generation of non-equilibrium by the dc-bias required to records the
$IV$ characteristics.

In this Letter, we present an investigation of the current phase
relation in long diffusive SNS junctions using Waldram's method
\cite{waldram,frolov} with a micro-Hall sensor \cite{geim} to detect
the magnetic flux. In thermodynamic equilibrium, we find
quantitative agreement with the quasi-classical theory. In
particular, the equilibrium CPR was found to be entirely sinusoidal
at higher temperatures. When the samples are irradiated with
microwaves with a photon energy comparable to
$\epsilon_\text{g}(\varphi)$, we could \emph{induce} a significant
contribution of higher harmonics in the CPR with an anomalous
$T$-dependence. Our observations can be understood in terms of a
non-equilibrium occupation of Andreev-levels in the microwave field,
and are important for the control of Andreev-quantum bits
\cite{petrashov}.

\begin{figure}[t]
\includegraphics[width=80mm]{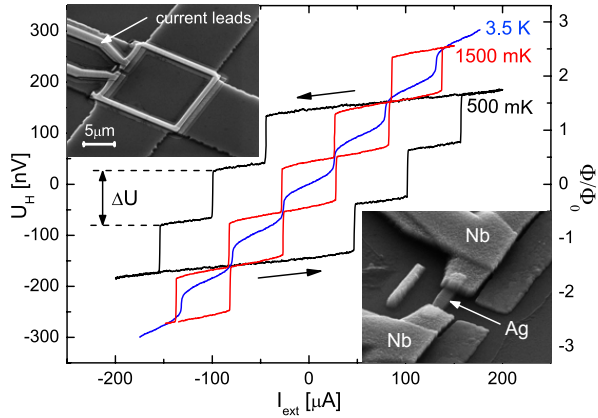}
\caption{(color online) The measured Hall voltage $U_\mathrm{H}$ as
a function of the loop current $I_\mathrm{ext}$.  For low
temperatures ($\beta>1$) the effective flux shows hysteretic
behavior, hence only the rising branch of the CPR is accessible.
Upper inset: SEM image of the Nb loop with an edge length of
10\,$\mu$m and an inductance $L\approx 40$\,pH fabricated on the
active area of a micro-Hall-cross. Lower inset: close-up of the
actual SNS junction. The thickness of the Ag (Nb) layer is 40\,nm
(150\,nm). While the spacing between the Nb electrodes is
$d=495\,$nm, the effective length of the Ag bridge is
$d_\text{eff}=640\,\text{nm}$ with a width of 210\,nm.}
\label{raw_data}
\end{figure}

To investigate the supercurrent response of our SNS-junctions to an
applied phase difference, we integrated the junctions into
superconducting loops. The gauge invariant phase difference
$\varphi$ across the junction is then related to the total magnetic
flux $\Phi$ via $\varphi=-2\pi\Phi/\Phi_0$, where $\Phi_0=h/2e$ is
the magnetic flux quantum \cite{Tinkham}. The samples were patterned
on top of the active area of micron-sized Hall-crosses (see insets
in Fig.~\ref{raw_data}) which have been structured into a
GaAs/AlGaAs semiconductor heterostructure containing a 2-dimensional
electron gas 190\,nm below the surface. With a mean free path of
$l=9$\,$\mu$m at 4\,K and an active area of $10 \times 10$\,$\mu
\mathrm{m}^2$ these Hall probes work close to the ballistic regime.
Depending on the bath temperature, which sets an upper limit to the
applied current through the Hall probe, $I_\mathrm{p}$, we obtain a
supercurrent sensitivity between 0.1 and 2\,$\mu$A, corresponding to
a flux sensitivity of $10^{-3}$ to $2\cdot10^{-2}$\,$\Phi_0$. The
fabrication process makes use of e-beam lithography and a shadow
evaporation technique based on a thermostable trilayer mask system.
We use a sacrificial layer of polyether sulfone (PES) as described
in the work of Dubos \emph{et al.} \cite{dubos2} and a mask layer of
silicon nitride (SiN) \cite{Gaass}. The insets to
Fig.~\ref{raw_data} show scanning electron micrographs of the loop
on top of the Hall cross, and a close-up of the junction,
respectively.

The magnetic flux is controlled by driving an external current
$I_\mathrm{ext}$ through the loop via current leads close to the SNS
contact (Fig.~\ref{raw_data}, upper inset). This has the advantage
that both the external flux $LI_\mathrm{ext}$ and the flux
$LI_\mathrm{S}$ induced by the junction are detected with the same
filling factor $\alpha$, which greatly facilitates the calibration
of the measured Hall voltage $U_\mathrm{H}$ in terms of $\Phi_0$:
\begin{equation}\label{flux}\Phi=\alpha
\Phi_\mathrm{H}=\alpha'U_\mathrm{H}=L\left(I_\mathrm{ext}-I_\mathrm{S}(2\pi\Phi/\Phi_0)\right).
\end{equation}
Here, $\Phi_\mathrm{H}$ is the flux enclosed by the active area $A$
of the Hall cross, $\alpha'=\alpha Ane/I_\mathrm{p}$,
$e=1.602\cdot^{-19}$, $n=2,26\cdot10^{15}$\,m$^{-2}$, and $L$ the
geometric inductance of the loop. Depending on the match between the
loop area and $A$, $\alpha$-values between 0.3 and 0.4 can be
achieved, limited by the finite distance between the 2DEG and the
loop.

Figure \ref{raw_data} shows the Hall voltage
$U_\mathrm{H}(I_\mathrm{ext})$ of a sample with $d=495$\,nm for
different temperatures. The magnetic response of the SNS loop is
superimposed on a linear background caused by $I_\mathrm{ext}$, the
slope of which can be used to determine $L$. Below a certain
temperature, when the parameter $\beta_\text{L}=2\pi
LI_\text{C}(T)/\Phi_0$ exceeds 1, the response is hysteretic with
distinct jumps when the critical current of the junction is
exceeded. In the limit $\beta_\text{L}\gg1$, the step height
corresponds to one flux quantum and can thus be used to determine
$\alpha'$. A plot of
$I_\mathrm{S}=I_\mathrm{ext}-\alpha'U_\mathrm{H}/L$ vs.
$\varphi=2\pi\alpha'U_\mathrm{H}/\Phi_0$ represents the desired CPR.

\begin{figure}[t]
\includegraphics[width=86mm]{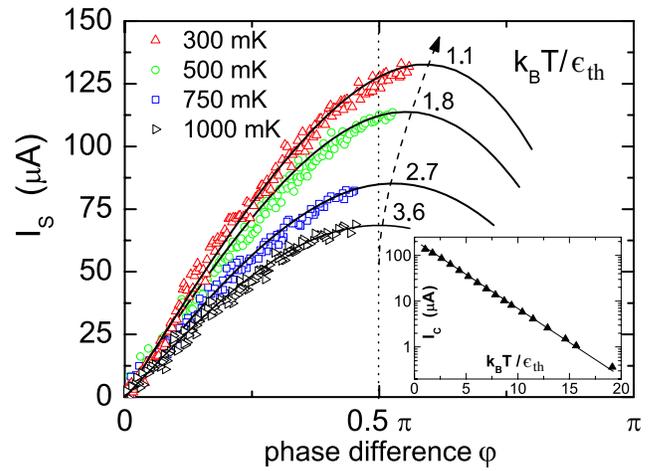}
\caption{(color online) CPR for low temperatures down to
$k_\mathrm{B}T\approx \epsilon_\text{th}$. For $T\lesssim 500$\,mK a
clear deviation from the sinusoidal Josephson relation is observed.
The solid lines correspond to the predictions of the quasiclassical
theory, with all parameters determined independently. Inset: The
critical current as a function of temperature. From the fit of the
data to the quasiclassical theory (solid line) we determine the
value $D=0.014\,$m$^2$/s of the diffusion constant in the Ag
bridge.}\label{CPR_eq}
\end{figure}

The Thouless energy $\epsilon_\text{th}$ of our junctions is
determined by the effective length $d_\mathrm{eff}>d$ of the normal
conducting barrier, which accounts for the fact that the Andreev
reflections at the SN interface occur randomly within the overlap
length $s$ between Nb and Ag. As in earlier work \cite{dubos3}, we
have chosen $d_\mathrm{eff}=d+s$. The gap parameter
$\Delta=1.3\,$meV of our Nb-films has been extracted from its
measured $T_c\simeq8.5\;$K. Fig.~\ref{CPR_eq} shows the measured CPR
for several temperatures below 1~K. While at 1~K the CPR is still a
perfect sine, at lower temperatures clear deviations from the
sinusoidal shape are observed. The inset to Fig.~\ref{CPR_eq} shows
 $I_\mathrm{C}(T)$, extracted from the maximum supercurrent \cite{note0},
 which agrees very well with the quasi-classical theory
\cite{AZaikin,heikk} using $D= 0.014\;$m$^2$/s as the only fitting
parameter (solid line in inset). This value of $D$ corresponds to a
$\epsilon_\text{th}=24\;\mu$eV for this sample
($d_\text{eff}=640$~nm). With all parameters fixed, we are now
prepared for a quantitative comparison of the measured CPR with the
theoretical predictions \cite{heikk}, represented by the solid lines
in Fig.~\ref{CPR_eq}. Within our experimental resolution the
agreement is excellent \cite{note1}. In particular, the result for
our lowest accessible temperature of
$300\,\mathrm{mK}=1.1\,\epsilon_\text{th}/k_B$ almost coincides with
the theoretical predictions for this value of
$\Delta/\epsilon_\text{th}=58$ in the limit $T\rightarrow 0$, where
the degree of anharmonicity is largest.

\begin{figure}[t]
\includegraphics[width=80mm]{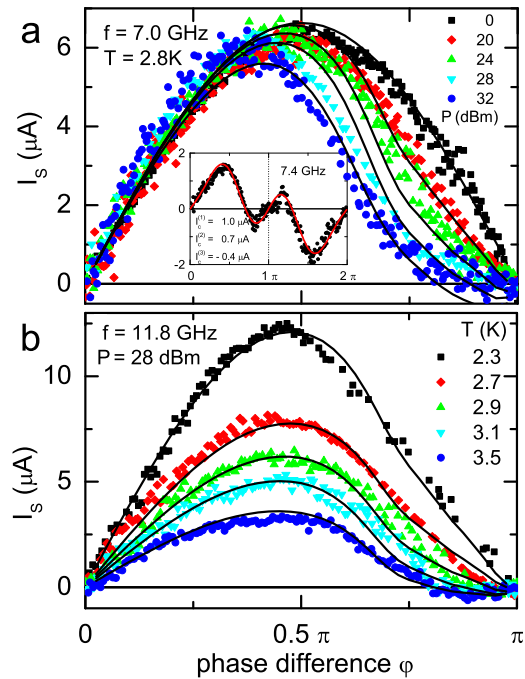}
\caption{(color online) a) Measured CPR under irradiation with
microwaves at $T=2.8\,$K and $f=7.0$\,GHz.  b) Measured CPR for
different temperatures at $f=11.8$\,GHz. Solid lines represent the
best fit to our model.  Inset: Similar data at 7.4~GHz (near a
cavity resonance). The solid line is a Fourier-expansion with the
three coefficients listed.}\label{CPR_rf}
\end{figure}

We now turn to the main subject of the present paper -- the effect
of microwave irradiation on the CPR in the frequency range between 3
and 12\,GHz. The rf-power is applied at room temperature and reduced
by several attenuators at 4\,K, 1\,K and 100\,mK (in total $-60$\,dB
including cable losses) to ensure proper thermalization of the
rf-cable, which is terminated with a small loop in the vicinity of
the sample.  Figure \ref{CPR_rf}a shows the CPR measured at 2.85~K
for different levels of the externally applied power $P$. These data
are taken away from a cavity resonance of the sample chamber. At
this temperature, the equilibrium CPR for $P=0$ is nearly sinusoidal
and the critical current is small enough ($\beta_\text{L}<1$) to
access the full CPR. It is clearly seen that with increasing power
the maximum supercurrent is reached for \emph{lower}
$\varphi$-values, $\varphi_\mathrm{max}<\pi$/2, in contrast to the
behavior at low temperatures (see Fig.~\ref{CPR_eq}). The effect is
most prominent for phase differences close to $\pi$, where a
pronounced suppression of the supercurrent occurs. For
$\varphi<\varphi_\mathrm{max}$, the supercurrent even appears to be
slightly enhanced compared to $P=0$. At certain frequencies, which
we attribute to resonances of our sample chamber, we found that
$I_\mathrm{C}^{(1)}$ and $I_\mathrm{C}^{(2)}$ can even be comparable
(see inset in Fig.\,\ref{CPR_rf}a). This can even lead to an {\it
inversion} of the supercurrent direction close to $\varphi\lesssim
\pi$. In Fig.\,\ref{CPR_rf}b the temperature dependence of the
effect is investigated at $P=28\,$dBm. The deviation from the
sinusoidal form is moderate at 2.3~K and becomes more pronounced as
$T$ is raised to 3.5~K. Again the effects are most pronounced around
$\varphi \simeq \pi$. At higher temperatures, the supercurrent
becomes too small to be detected with our method. We have obtained
qualitatively similar results on a shorter, but otherwise similar
junction with $\Delta/\epsilon_\text{th}=39$.

What is the origin of the microwave induced suppression of the
supercurrent around $\varphi\simeq \pi$? The rf-field is expected to
excite small oscillations of the magnetic flux enclosed by the loop
\cite{note2}. One possibility is that these oscillations lead to the
induction of an ac-voltage $V_\mathrm{ac}$ across the junction and
modify the distribution function $f(\epsilon)$ of Andreev states
within the junction. One would then expect sidebands in
$f(\epsilon)$ of width $\hbar\omega_\mathrm{rf}$ \cite{carlos}.
Indeed, such a mechanism can lead to a distortion of the CPR
qualitatively similar to our observations. However, it disagrees
with our data in that the $I_\mathrm{C}^{(j)}(T)$ should again
vanish exponentially for $j>1$ and $T\gtrsim
\epsilon_\text{th}/k_B$.

Another possibility are single-photon excitations of
quasi\-particles to ABS carrying supercurrent in the opposite
direction. Such excitations become possible when the minigap goes to
zero near $\varphi\simeq\pi$. To model this effect, we start, as in
the equilibrium case, from Eq.~\ref{SC_spect}, assuming that one can
use the {\em equilibrium} supercurrent spectral density ${\rm
Im}\,j_\mathrm{S}(\epsilon,\varphi)$ with a {\em nonequilibrium}
distribution function $f(\epsilon)$.

A kinetic equation for $f(\epsilon)$ in dirty superconductors under
microwave irraduation was formulated first by Eliashberg
\cite{Eliashberg}. It can be represented in the form
\begin{align}
\label{Eliashberg-2} \nonumber
  P_\omega\Big[A(\epsilon) & \Big(f(\epsilon-\omega)-f(\epsilon)\Big)
  +B(\epsilon)\Big(f(\epsilon+\omega)-f(\epsilon)\Big) \\
&
+C(\epsilon)\Big(1-f(\omega-\epsilon)-f(\epsilon)\Big)\Big]=I[f(\epsilon)],
\end{align}
where the coefficient
$P_\omega=D|\delta\varphi_\omega/d_\text{eff}|^2$ is proportional to
the rf intensity. The phase oscillations $\delta \varphi_\omega$
represent the ac-response of the loop when excited by the microwave
field. The coefficients $A(\epsilon)$, $B(\epsilon)$, and
$C(\epsilon)$ can be expressed through the Green's functions of the
system \cite{Eliashberg}. We have approximated the Green's functions
of our proximity superconductor by the standard BCS form and
replaced $\Delta$ with the minigap $\epsilon_g(\varphi)$.
$I[f(\epsilon)]$ is the scattering integral including
electron-electron (ee) and electron-phonon (ep) interactions. The
first two terms $\propto P_\omega$ describe {\em scattering} of
quasiparticles by photons and the third term {\em pair-breaking},
i.e., the creation of electron-hole pairs by single-photon
absorption.

Assuming that the ee-scattering is much stronger than the
ep-scattering, we approximate $f(\epsilon)$ by Fermi functions
$f^*(\epsilon)$ with symmetrically shifted chemical potentials
$\mu^{*}$ and $-\mu^*$ \cite{Tinkham} for the microwave excited
electron- and hole-like quasiparticles, respectively. Since the
excitation of quasiparticles by photons creates no charge imbalance,
the nonequilibrium distribution function $f^*$ is particle-hole
symmetric. Then $f^*$ and $\mu^*$ can be determined from
Eq.~\ref{Eliashberg-2} within a relaxation time approximation,
rather than numerically solving the inhomogeneous kinetic equation.
We neglect the $T$-dependence of the ep-scattering time
$\tau_\mathrm{ep}$ in the limited $T$-interval under consideration.
After integration of Eq.~\ref{Eliashberg-2} over energy the terms
containing $A(\epsilon)$ and $B(\epsilon)$ vanish, because they do
not change the total number of excitations, and we obtain
\begin{equation}\label{balance}
P_\omega\int_0^\infty C(\epsilon)
  \Big(1-f^*(\omega-\epsilon)-f^*(\epsilon)\Big)d\epsilon=\frac{\Delta n}{\tau_\mathrm{ep}},
\end{equation}
where $\Delta
n=\int_0^{\infty}\left(f^*(\epsilon)-f^0(\epsilon)\right)d\epsilon$
and $f^0(\epsilon)$ is the equilibrium Fermi distribution function.

\begin{figure}[t]
\includegraphics[width=90mm]{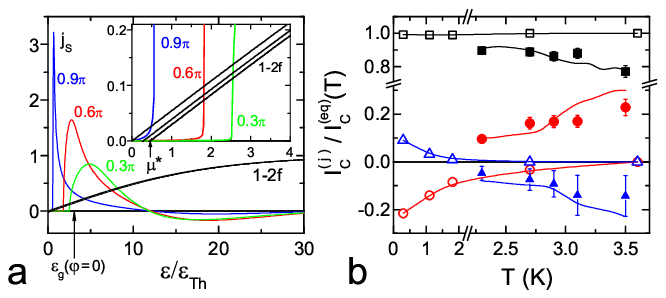}
\caption{(color) a) Spectral supercurrent $j_\text{S}$ for
$\varphi/\pi=0.3$, 0.6 and 0.9 together with $1-2f^*$ vs.~$\epsilon$
at 2.8~K. Inset: Zoom for $\epsilon/\epsilon_\text{th}\leq 3$. Black
lines: $1-2f^*$ for $P=0$, 24 dBm and 32 dBm (from top to bottom).
The point $1-2f=0$ defines the chemical potential $\mu^*$ for
electron-like quasi-particles. b) Normalized values of
$I_\mathrm{C}^\mathrm{(1)}$ ($\blacksquare,\square$),
$I_\mathrm{C}^\mathrm{(2)}$
(\color[rgb]{1,0,0}\ding{108},\textsf{O}\color[rgb]{0,0,0} ) and
$I_\mathrm{C}^\mathrm{(3)}$
(\color[rgb]{0,0,1}\ding{115},$\triangle$\color[rgb]{0,0,0}) for the
measured CPR without rf-irradiation (open symbols) and for 11.8\,GHz
and $P=28\,$dBm (filled symbols) as a function of temperature. The
lines are the theoretical curves corresponding to $f^0$ and
$f^*$.}\label{fourier}
\end{figure}

In Fig.~\ref{fourier}a we plot $j_\text{S}(\epsilon)$ together with
$1-2f^*(\epsilon)$. Despite the resulting small values
$\mu^*\lesssim0.5\,\epsilon_\text{th}$, we have a noticable effect
on the total supercurrent $I_\text{S}(\varphi)$ around
$\varphi\simeq\pi$. The increase of $\mu^*$ cuts off the sharp
positive peak of $j_\text{S}$ at $\epsilon_\text{g}(\varphi)$ and
enhances the relative weight of the negative part of
$j_\text{S}(\epsilon)$ at higher $\epsilon$ in Eq.~\ref{SC_spect}.
This can explain the observed reversal of $I_\text{S}(\varphi)$. We
express
$P_\omega\tau_{ph}=\eta(\omega)\exp\left(P[\text{dBm}]/10\right)$ in
terms of a single parameter $\eta$, which also contains the
$\omega$-dependent coupling between the sample loop and the antenna.
$P$ is the externally applied microwave power. The results of our
calculations are displayed as solid lines in Fig.~3 with
$\eta(\omega)$ kept fixed to $0.001$ for $f=7.0\,$GHz and $0.005$
for $11.8\,$GHz. All other parameters are identical to that of
Fig.~\ref{CPR_eq}. The agreement achieved is quite satisfactory.

In  Fig.~\ref{fourier}b we show the Fourier coefficients of the
measured CPR together with those extracted from the theory up to the
3$^\mathrm{rd}$ harmonics. For a better comparison, the data are
normalized with respect to the equilibrium critical current
$I_\mathrm{C}^\mathrm{(eq)}(T)$. In thermodynamic equilibrium (open
symbols), the sign of the Fourier coefficients alternates and they
decay monotonically with temperature. Under rf-irradiation (full
symbols), $I_\mathrm{C}^\mathrm{(1)}$ slightly decreases with
microwave power, which may result from slight electron heating by
the microwaves. In contrast, $I_\mathrm{C}^\mathrm{(2)}$ and
$I_\mathrm{C}^\mathrm{(3)}$ are much larger, {\it increase} with
$T$, and have opposite sign when compared to the case without rf
irradiation. The experimental data agree rather well with the
theoretical values. Our observations resemble very much the peculiar
temperature dependence of the amplitude of the subharmonic Shapiro
steps observed in Refs.~\cite{lehnert,dubos}.

In conclusion, we have experimentally verified long-standing
theoretical predictions on the current-phase relation in diffusive
Josephson junctions with highly transparent interfaces. We have
presented strong evidence that microwave-irradiation of the junction
induces higher harmonics in the current-phase relation. This effect
can be understood in terms of a non-equilibrium occupation of
Andreev bound states carrying the supercurrent through the junction.
Our results call for caution in the use of subharmonic Shapiro steps
as a method for the determination of the current-phase relation.

We thank E.~Scheer for support in instrumentation and
T.~Heikkil\"{a} and J.~C.~Cuevas for providing codes for a numerical
evaluation of the spectral supercurrent and for stimulating
discussions. This work has been supported by the DFG (STR 438-2 and
GRK 638).


\begin{thebibliography}{99}

\bibitem{kulik}
I.O.~Kulik, Zh. Eksp.~Teor.~Fiz. \textbf{57}, 1745 (1969)
[Sov.~Phys.~JETP \textbf{30}, 944 (1970)].
\bibitem{ilichev}
For a review see, e.g., A.~Golubov, M.~Kupriyanov, and E.~Il'ichev,
Rev. Mod. Phys.~{\bf 76}, 411 (2004) and the references therein.
\bibitem{heikk} T.T.~Heikkil\"{a}, J.~S\"{a}rkk\"{a}, F.~K. Wilhelm, Phys. Rev. B {\bf 66}, 184513 (2002).
\bibitem{waldram}
J.R.~Waldram and J.M.~Lumley, Rev.~Appl.~Phys., \textbf{10}, 7
(1975).
\bibitem{frolov}
S.M.~Frolov, {\it et al.}, Phys. Rev. B \textbf{70}, 144505 (2004).
\bibitem{lehnert}
K.W.~Lehnert {\it et al.}, Phys. Rev. Lett. \textbf{82}, 1265
(1999).
\bibitem{dubos}  P.~Dubos, H.~Courtois, O.~Buisson, B.~Pannetier,
                Phys.~Rev.~Lett.~{\bf 87}, 206801 (2001).
\bibitem{petrashov}
J.~Skoldberg {\it et al.}, Phys.~Rev.~Lett.~{\bf 101},
087002 (2008).
\bibitem{geim}A.K.~Geim {\it et al.}, Appl.~Phys.~Lett.~{\bf 71}, 2379 (1997).
\bibitem{Tinkham}{\it Introduction to Superconductivity}, M.~Tinkham, McGraw-Hill, New York, 2nd ed.
(1996).
\bibitem{dubos2}  P.~Dubos {\it et al.}, J.~Vac.~Sci.~Technol.~B {\bf 18}, 122 (2000).
\bibitem{Gaass}M.~Gaass {\it et al.}, Phys.~Rev.~ B {\bf 77}, 024506 (2008).
\bibitem{dubos3}P.~Dubos {\it et al.},
Phys.~Rev.~B \textbf{63}, 064502 (2001).
\bibitem{note0}
Premature switching caused by thermal or quantum fluctuations can be
neglected in our junctions.
\bibitem{AZaikin}
A.D.~Zaikin and G.F.~Zharkov, Fiz.~Nizk.~Temp.~\textbf{7}, 375
(1981); [Sov.~J.~Low Temp.~Phys.~\textbf{7}, 184 (1981)].
\bibitem{note1}
Neglecting the geometric inductance of the junction results in an
underestimation of $I_\text{C}$, which is less than 2\% in our
samples.
\bibitem{note2}
These oscillations are strongly damped, because of the low ($\simeq$
1 $\Omega$) quasi-particle resistance of the junctions.
\bibitem{carlos}J.C.~Cuevas, private communication.
\bibitem{Eliashberg} G.M.~Eliashberg and B.I.~Ivlev in {\it Nonequilibrium Superconductivity},
 Edts.~D.N.~Langenberg and A.I.~Larkin (North Holland 1986), p.~211.


\end{thebibliography}
\end{document}